\documentclass[runningheads]{cl2emult}

\usepackage{makeidx}  
\usepackage{graphicx} 
\usepackage{subeqnar} 
\usepackage{multicol} 
\usepackage{cropmark} 
\usepackage{eso}      
\makeindex            

\def\ie{{i.e.,~}}
\def\eg{{e.g.,~}}
\def\etal{{et al.~}}
\def\spose#1{\hbox to 0pt{#1\hss}}
\def\simlt{\mathrel{\spose{\lower 3pt\hbox{$\mathchar"218$}}
     \raise 2.0pt\hbox{$\mathchar"13C$}}}
\def\simgt{\mathrel{\spose{\lower 3pt\hbox{$\mathchar"218$}}
     \raise 2.0pt\hbox{$\mathchar"13E$}}}
\def\plotfiddle#1#2#3#4#5#6#7{\centering \leavevmode
\vbox to#2{\rule{0pt}{#2}}
\includegraphics{#1}}

\begin{document}
\title*{First Results from the SPICES Survey}
\toctitle{First Results from the SPICES Survey}
%
%
\titlerunning{First Results from SPICES}
%
\author{Daniel Stern\inst{1}
\and Andrew Connolly\inst{2}
\and Peter Eisenhardt\inst{1}
\and Richard Elston\inst{3}
\and Brad Holden\inst{4}
\and Piero Rosati\inst{5}
\and S.~Adam Stanford\inst{4}
\and Hyron Spinrad\inst{6}
\and Paolo Tozzi\inst{7}
\and K.~L.~Wu\inst{3}}
\authorrunning{Daniel Stern et al.}
%
%
\institute{Jet Propulsion Laboratory, California Institute of
	Technology, MS~169-327, Pasadena, CA 91109, USA
\and Department of Physics and Astronomy, University of Pittsburgh,
	Pittsburgh, PA 15260, USA
\and Department of Astronomy, The University of Florida, P.O.~Box~112055,
	Gainesville, FL 32611, USA
\and Physics Department, University of California at Davis,
	Davis, CA 95616, USA and Institute of Geophysics and
	Planetary Physics, Lawrence Livermore National Laboratory,
	Livermore, CA 94550, USA
\and ESO
\and Department of Astronomy, Univeristy of California at Berkeley, 
	Berkeley, CA 94720, USA
\and Osservatorio Astronomico di Trieste, via G.B. Tiepolo~11,
	I-34131, Trieste, Italy}

\maketitle              

\begin{abstract}

We present first results from SPICES, the Spectroscopic, Photometric,
Infrared-Chosen Extragalactic Survey.  SPICES is comprised of four
$\approx 30$ arcmin$^2$ high Galactic latitude fields with deep
$BRIzJK_s$ imaging reaching depths of $\approx 25$ mag (AB) in the
optical and $\approx 23$ mag (AB) in the near-infrared.  To date we
have 626 spectroscopic redshifts for infrared-selected SPICES sources
with $K_s < 20$ (Vega).  The project is poised to address galaxy
formation and evolution to redshift $z \approx 2$.  We discuss initial
results from the survey, including the surface density of extremely red
objects and the fraction of infrared sources at $z > 1$.  One of the
SPICES fields has been the target of a deep 190~ksec {\it Chandra}
exposure; we discuss initial results from analysis of that data set.
Finally, we briefly discuss a successful campaign to identify
high-redshift sources in the SPICES fields.

\end{abstract}

\section{Introduction }

The past few years have been a watershed in our ability to directly
observe galaxy evolution.  Deep field surveys such as the Canada-France
Redshift Survey (CFRS -- Lilly \etal 1995) and color-selected field
samples such as that of Steidel \etal (1996, 1999) have provided
critical information on the evolution of field galaxies.  Madau \etal
(1996) integrated the results at $z < 5$ into a coherent picture of the
star formation history of the Universe, suggesting that the global star
formation rate peaked between $z = 1$ and $2$.  Since then, recognition
of the importance of both dust and cosmic variance has changed the
steep decline in the cosmic star formation rate inferred at $z > 2$
into a flat plateau for $1 \simlt z \simlt 4$ (Steidel \etal 1999).
Cowie \etal (1999) also show a more gradual rise at $z < 1$ than
initially inferred by the CFRS. 

There remain four substantial caveats regarding these findings.  First,
the number of spectroscopically measured redshifts between $z=1$ and
$2$ is small.  Second, since the UV dropout technique used to identify
the $z \simgt 3$ population requires them to be UV bright, it is
possible that a substantial amount of star-forming activity in dusty
systems has been overlooked.  Third, redshift surveys from which cosmic
star formation rates are measured must be of sufficient depth and
wavelength coverage that star formation indicators (\eg, $M$(2800 \AA))
can be measured with limited extrapolation over wide redshift
intervals.  Finally, small area surveys, such as the HDF, are
vulnerable to perturbations from large scale structure.

Infrared-selected surveys provide a powerful tool for addressing these
issues (\eg see Dickinson, these proceedings).  Among the benefits,
infrared $k$-corrections are small and relatively independent of galaxy
type, age, and redshift.  Since the long-wavelength light of galaxies
is dominated by lower mass stars rather than short-lived high-mass
stars, infrared luminosities track galaxy mass, thereby providing a more
direct comparison to theories of galaxy formation without relying on
the poorly-understood physics of star formation.  Infrared light is
also less vulnerable to dust absorption.

On the negative side, since spectroscopy is primarily performed at
optical wavelengths, infrared-selected samples are challenging to
follow-up.  Also, since evolved stars become important contributors to
the long-wavelength flux of a galaxy, poorly-understood phases of
stellar evolution can make interpretation of broad-band colors
ambiguous (\eg Spinrad \etal 1997).  Finally, since infrared-surveys do
not select for young stars, they are suboptimal for studying the cosmic
star-formation history, though they provide a natural basis for
studying the mass-aggregation history.

\section{First Results from SPICES }

We present the SPICES survey (Eisenhardt \etal 2001, in prep.), a deep
$BRIzJK_s$ imaging and spectroscopic survey covering over 100 arcmin$^2$
spread over four fields.  Table~\ref{Tab1} lists the Vega magnitude
3$\sigma$ depths in 3'' diameter apertures for the imaging.  The
relatively large area mitigates the effects of large-scale structure
while the $K$-band depth is more than sufficient to detect $L^\ast$
galaxies to $z = 2$.  The area and depth are a significant improvement
over several recent surveys (\eg Cowie \etal 1996), but are modest
compared to several programs currently in production mode (\eg Cimatti,
these proceedings; McCarthy, these
proceedings).  An important strength of SPICES is the spectroscopic
program:  we currently have 626 spectroscopic redshifts of $K < 20$
sources selected from the sample, approximately one-third of the
complete $K < 20$ sample (see Figure~1).  These spectroscopic redshifts
are being used to directly construct an eigenbasis of galaxy spectral
energy distributions with which to determine photometric redshifts for
the complete sample (see Budavari \etal 2000).  Wu \etal
(these proceedings) discusses {\it HST} imaging of one of the SPICES
fields.  Here we discuss two initial results from the survey.

\begin{table}
\centering
\caption{Depth of SPICES Imaging (Vega magnitudes)}
\renewcommand{\arraystretch}{1.4}
\setlength\tabcolsep{5pt}
\begin{tabular}{lcccccc}
\hline\noalign{\smallskip}
filter & $B$ & $R$ & $I$ & $z$ & $J$ & $K_s$ \\
\noalign{\smallskip}
\hline
\noalign{\smallskip}
3$\sigma$ depth & 27 & 25.5 & 25.0 & 24.2 & 23.0 & 21.3 \\
\hline
\end{tabular}
\label{Tab1}
\end{table}

\begin{figure}[t]
\plotfiddle{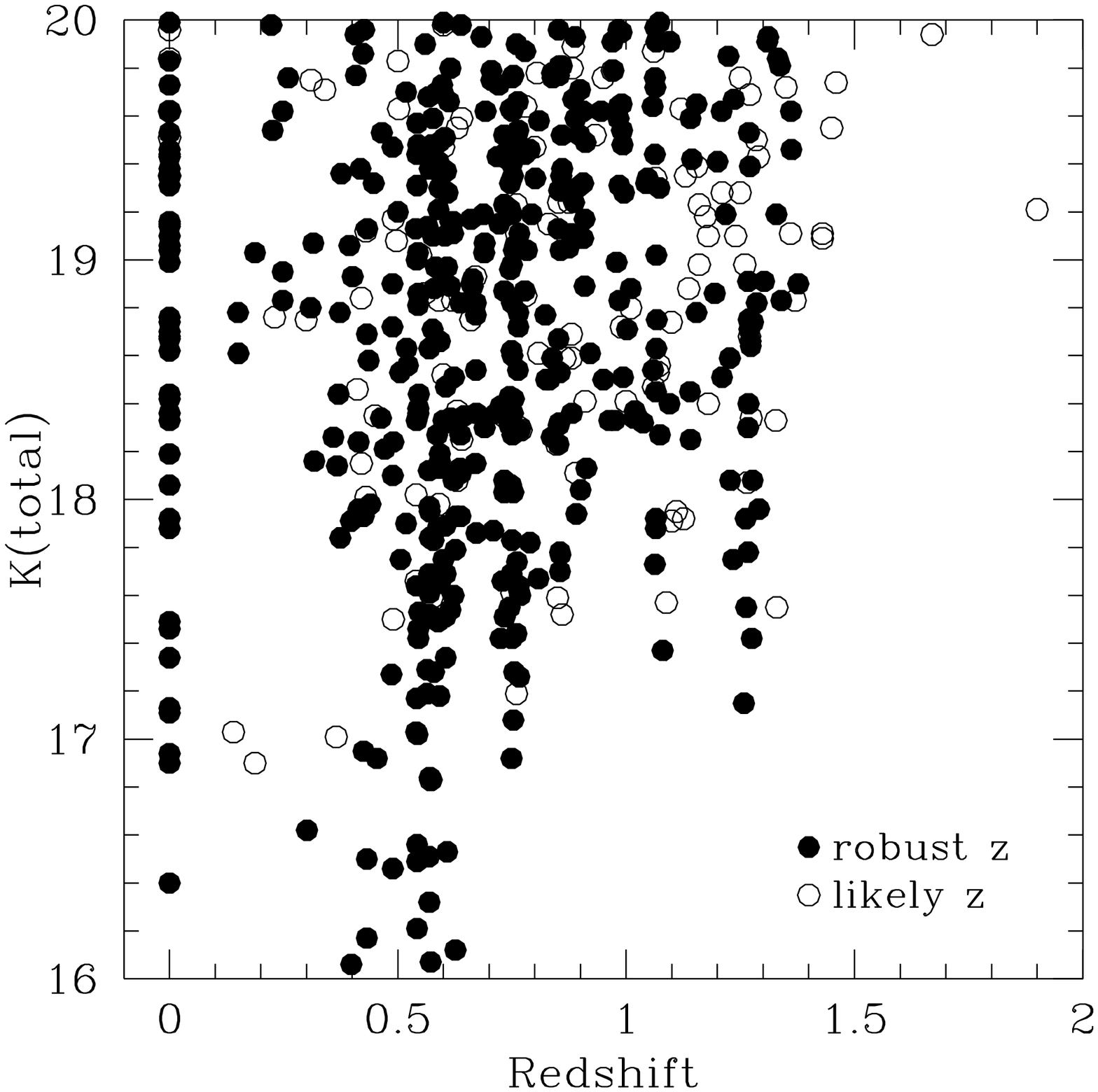}{0.8in}{0}{25}{25}{-150}{-110}
\plotfiddle{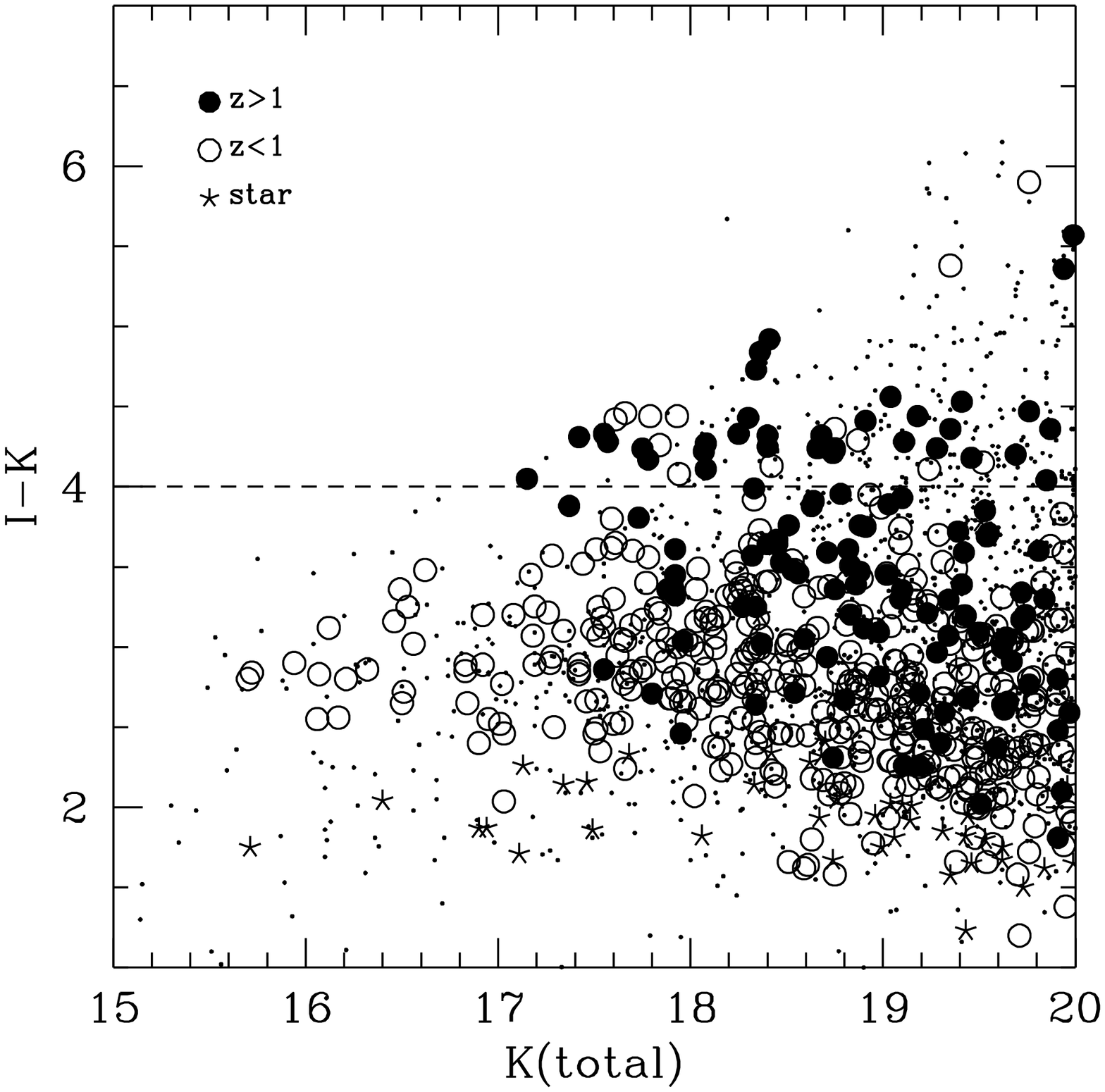}{0.8in}{0}{25}{25}{10}{-41}

\caption{Status of the current survey.  We now have 626 spectroscopic
redshifts over the four fields which comprise the survey.  {\bf Left:}
Location of the spectroscopic redshifts in $K-z$ space.  Filled-in
circles indicate unambiguous redshifts, while open circles indicate
likely redshifts.  Note the vertical stripes corresponding to large
scale structures at several redshifts between $z \approx 0.6$ and $z
\approx 1.3$.  {\bf Right:}  Location of SPICES sources in
color-magnitude space.  Dots correspond to objects lacking spectroscopic
redshifts.  Symbols indicate redshift for spectroscopic targets.  Note
that $I-K > 4$ is a robust indicator of $z > 1$.}

\end{figure}

\noindent \underline{\it The Surface Density of ERO's:}  Extremely red
objects (ERO's) are an intriguing class of extragalactic object, likely
associated with $z \simgt 1 $ galaxies (\eg Cimatti, these
proceedings).  We find that the surface density of these sources is
elevated in the SPICES fields relative to some of the surface densities
reported previously in the literature.  For $I - K < 4, 19 < K < 20$,
we find a surface density of 1.4 ERO's arcmin$^{-2}$, with a range in
this value of 1.3 to 1.9 across the four fields.  For the same
magnitude range and color criterion, Barger \etal (2000) find a surface
density of $0.2 \pm 0.1$ ERO's arcmin$^{-2}$ over a field of view of
61.8 arcmin$^2$ while McCracken \etal (2000) find a surface density of
ERO's in the Herschel Deep Field of $\approx 0.5 \pm 0.1$ arcmin$^{-2}$
over a 47.2 arcmin$^2$ field.  Similarly, if we consider ERO's defined
as $K < 19$ sources with $R - K > 6$, the SPICES fields have 0.13 ERO's
arcmin$^{-2}$ with a range of $0.06 - 0.32$ ERO's arcmin$^{-2}$ across
the four fields.  Using the same definition, the CADIS survey finds
$0.039 \pm 0.016$ ERO's arcmin$^{-2}$ across a 154 arcmin$^2$ field
(Thompson \etal 1999) while Daddi \etal (2000) find 0.07 ERO's
arcmin$^{-2}$ across a 447.5 arcmin$^2$ field with strong clustering
reported.

What is the source of this discrepancy?  One possibility is that the
depth and area of the SPICES imaging are significantly improved over
many of the surveys mentioned above:  $K = 20$ is a 10$\sigma$
detection in the SPICES survey.  Another possibility is large scale
structure.  Though the SPICES fields cover $> 100$ arcmin$^2$, larger
than several of the above surveys, fluctuations in the ERO surface density on
these scales have been reported by more recent larger area deep
infrared surveys (\eg Daddi \etal 2000; Cimatti, these proceedings;
McCarthy, these proceedings).  Indeed, one of the SPICES fields (the
Lynx field:  $08^{hr}48^{min}, +44^o54'$) has a higher surface density
of red objects than the other three fields.  Keck/LRIS spectroscopy and
has subsequently identified many of these red sources with galaxies in
two X-ray emitting clusters at $z \sim 1.27$ (Stanford \etal 1997;
Rosati \etal 1998).

\medskip

\noindent \underline{\it The $z > 1$ Fraction:}  The $K$-band
luminosity function (KLF) at $z = 1$ offers a powerful constraint on
theories of galaxy formation.  Since the $K$-band light tracks mass
better than ultraviolet/optical light, the KLF is more directly
comparable to theories of the collapse and merging of galaxies.
Kauffmann \& Charlot (KC98; 1998) show that pure luminosity evolution
(PLE) models, \ie models in which galaxies collapse monolithically at
high redshift with little subsequent merging activity, predict that
many massive galaxies exist at $z = 1$:  $\approx 54$\% of an
infrared-selected field galaxy sample with $18 < K < 19$ should be at
$z > 1$.  Alternatively, their hierarchical model predicts only
$\approx 3$\% of $18 < K < 19$ field galaxies should be at $z > 1$.
Ignoring the SPICES field with the $z \sim 1.27$ clusters and another
field with very limited spectroscopy, we conservatively find that $>
17$\% of $18 < K < 19$ SPICES sources are at $z > 1$.  This assumes
that $\approx 67$\% of $K < 19$, $I - K > 4$ (\ie red) sources are at
$z > 1$, as our spectroscopic program shows thus far, and we only count
those $18 < K < 19$, $I - K < 4$ (\ie blue) sources already
spectroscopically confirmed to be at $z > 1$.  Early photometric
redshift analysis on these fields suggests a value $\approx 25$\% of
the $18 < K < 19$ being at $z > 1$.  These numbers show that neither
PLE nor the KC98 hierarchical model correctly predicts the $z \approx 1$
KLF, implying that substantial merging occurs at $z > 1$.

\section{X-Ray SPICE and High-$z$ SPICE}

The identification of two clusters at $z \sim 1.27$ and one cluster at
$z = 0.56$ in the Lynx SPICES fields has led to a deep, 190~ksec {\it
Chandra} map of the field.  Analysis of the diffuse high- and
low-redshift cluster X-ray emission are discussed in Stanford \etal
(2001, submitted) and Holden \etal (2001, in prep.), respectively.
Stern \etal (2001, in prep.) discusses X-ray background (XRB) results
from this data set.  We confirm results of recently published {\it
Chandra} studies (\eg Giacconi \etal 2001):  most of the $0.5 - 10$ keV
XRB is resolved into discrete sources; the fainter soft-band sources
have harder X-ray spectra, providing a coherent solution to the
long-standing `spectral paradox'; and $\approx 90$\% of the sources
have optical/near-infrared identifications in deep ground-based imaging.  A
preliminary spectroscopic program shows a mix of obvious AGN,
apparently normal galaxies, and, perhaps surprisingly, several X-ray
emitting stars, some with hard X-ray spectra.

We are also targeting the SPICES fields with very deep imaging in
$RIz$ to identify high-redshift sources using the Lyman break
technique.  This work has led to the discovery of a faint quasar at $z
= 5.50$ (Stern \etal 2000) and several high-redshift galaxies out to $z
= 4.99$.  Strong emission-line galaxies have also been identified
serendipitously during the SPICES spectroscopic campaign, the highest
redshift source being a likely $z = 5.17$ Ly$\alpha$ emitter with
$f_{\rm Ly\alpha} \approx 9 \times 10^{-17}$ erg cm$^{-2}$ s$^{-1}$.

\section{Conclusions }

We present first results from the SPICES survey, an infrared-selected
photometric and spectroscopic survey.  We find an elevated surface
density of ERO's compared to several recent deep, infrared surveys,
likely due to fluctuations in that quantity from large scale structure
at moderate redshifts.  Perhaps relatedly, we also find a large
fraction of infrared-bright ($K < 19$) galaxies residing at $z > 1$.  A
good measure of this quantity provides a powerful constraint on models
of galaxy formation.

\medskip

Portions of this were carried out at the Jet Propulsion Laboratory,
California Institute of Technology, under a contract with NASA.

\clearpage
\addcontentsline{toc}{section}{Index}
\flushbottom
\printindex

\end{document}